\documentclass[10pt,english,a4paper]{paper}
\pdfoutput=1
	
\usepackage[T1]{fontenc}
\usepackage[latin9]{inputenc}
\usepackage{amsthm}
\usepackage{amsmath}
\usepackage{geometry}
\usepackage{lmodern}
\usepackage{babel}
\usepackage{color}
\usepackage{graphicx}
\usepackage[longnamesfirst]{natbib}

\title{Balancing Type I Error and Power in Linear Mixed Models}
\author{Hannes Matuschek$^a$ \and  Reinhold Kliegl$^a$ \and Shravan Vasishth$^a$ \and Harald Baayen$^b$ \and Douglas Bates$^c$}
\institution{$^a$ University of Potsdam \and $^b$ University of T\"{u}bingen \and $^c$ University of Wisconsin-Madison }

\begin{document}

\maketitle

\begin{abstract}
Linear mixed-effects models have increasingly replaced mixed-model analyses of variance for statistical inference in factorial psycholinguistic experiments. Although LMMs have many advantages over ANOVA, like ANOVAs, setting them up for data analysis also requires some care. One simple option, when numerically possible, is to fit the full variance-covariance structure of random effects \citep[the maximal model; ][]{Barr2013}, presumably to keep Type I error down to the nominal $\alpha$ in the presence of random effects. Although it is true that fitting a model with only random intercepts may lead to higher Type I error, fitting a maximal model also has a cost: it can lead to a significant loss of power. We demonstrate this with simulations and suggest that for typical psychological and psycholinguistic data, 
higher power is achieved without inflating Type I error rate if a model selection criterion is used to select a random effect structure that is supported by the data.
\end{abstract}

\section{Introduction}

During the last ten years, linear mixed-effects models \citep[LMMs, e.g., ][]{pinheirobates,Demidenko2004} have increasingly replaced mixed-model analyses of variance (ANOVAs) for statistical inference in factorial psycholinguistic experiments.  The main reason for this development is that LMMs have a number of advantages over ANOVAs.  From a pragmatic perspective, the most prominent one is that a single LMM can replace two separate ANOVAs with subjects ($F_1$ ANOVA) and items ($F_2$ ANOVA) as random factors, which does away with ambiguities of interpretation when effects are significant in only one of the two default ANOVAs. Other advantages are, for example, better preservation of statistical power in the presence of missing data \citep{baayen2008analyzing,pinheirobates} and options for simultanenous analyses of experimental effects and individual (or item) differences associated with them \citep{Gonzalez2014,KlieglEtAl2011}.  

The advantages of LMMs over ANOVAs come at a cost; setting up an LMM is not as straightforward as running an ANOVA. In response to this greater complexity of LMMs, \citet{Barr2013} proposed that LMMs with the most complex random-effect structure should be implemented for the analysis of factorial experiments as long as they converge. Here, we show on the basis of simulations, that this proposal may imply a significant loss in power whereas a  parsimonious mixed model \citep{Bates2015} containing only variance components supported by the data improves the balance between Type I error and power. To this end, we briefly review the core difficulty with LMM specification. 
 
An LMM requires not only the specification of the usual main effects and interactions of the experimental design (i.e., fixed effects), but also of variance components and correlation parameters associated with the random factors, usually subjects and items. In principle, aside from estimating variance components for the mean response of each random factor (i.e., random intercepts), the model estimates variance components for each within-subject and each within-item main effect and interaction term (i.e., random slopes) as well as associated correlations parameters. 

Moreover, again in principle, there is no, or only very rarely, reason to assume that some of these variance components or correlation parameters are truly zero, because there are no natural constants in fields like psychology or psycholinguistics. Unfortunately, even if the parameters are not zero, it is unlikely that all of them are sufficiently large to be reliably detected against the background of noise in the data (i.e., the residual variance). In part, this may also reflect limitations of optimizers in the available computer programs. As a consequence, the program may not converge to a stable solution; the model is overparameterized or degenerate relative to the information in the data. In this case, everybody agrees that it is up to the scientist---possibly with some help by the computer program---to reduce such an overparameterized model to one that is supported by the data, usually by fixing some of the small variance components or correlation parameters to zero \citep[for example, see discussion in][p.\ 276]{Barr2013}.\footnote{Because maximal models will usually converge in a Bayesian setting, an alternative could be to go the Bayesian route and fit models using probabilistic programming languages such as Stan \citep{stan-manual:2016}. \citet{SorensenHohensteinVasishthTutorial} and \citet{NicenboimVasishthStatMeth} provide detailed tutorials on how to fit (maximal) Bayesian LMMs, and \citet{Bates2015} shows examples of how maximal and non-maximal models can be fit using Stan. Whether it makes sense to fit overparameterized maximal models even in a Bayesian framework is a different question.}
 
Fortunately, with enough data for every subject and every item, the programs may converge and may provide what looks like an acceptable solution. Obviously, in such a model many of the model parameters may still be very close to zero and consequently removing such parameters from the model will probably not significantly decrease the goodness of fit of the model. How models are to be selected is itself a highly active field of research, covering all fields of advanced multivariate statistics (i.e., it is not an LMM-specific problem),  and, at this point in time, there is no perfect solution, but there are accepted ones \citep[e.g.,][]{Vandekerckhove2015}. Such reduction in model complexity leads to a simpler model than the maximal model. However, many psychologists do not want to engage in and possibly have to defend their model selection. And indeed, as long as the maximal model were to yield the same estimates as a justifiably reduced parsimonious model, there actually would be no need for model reduction. Most importantly, by fitting the maximal model, one avoids the risk of falsely removing a variance component which may lead to anti-conservative test statistics for fixed effects \cite[e.g.,][]{Barr2013, Schielzeth2009}. Similarly, failure to model correlations may increase the Type I error rate of the associated fixed effects above the specified significance criterion.%
\footnote{The significance level, usually 5\%, should guarantee a rate of 5\% false positives in the long run.} In other words, the maximal model protects a researcher from falsely reporting a spurious fixed effect that originates from a misspecification of the  unknown \emph{true} random-effect structure.

It was against this background that \citet{Barr2013} concluded on the basis of simulations that maximal LMMs yield appropriate test statistics for fixed effects. We paraphrase their recommendation as follows:

\begin{quotation}
[T]he maximal random effect structure should be fitted to the data. This includes a variance component for subject-related and item-related intercepts, for every within-subject and within-item fixed effect, and in the ideal case even all possible correlations between these random effects. The random effect structure should be reduced \emph{if and only if} the maximal model does not converge.  
\end{quotation}

In many ways, this is a surprising recommendation because it provides a very simple solution to a problem that is quite pervasive in many types of statistical analyses. But could there be a cost to potentially overfitting an LMM?\footnote{In \citet{Barr2013}, the authors already demonstrated that cost for the so-called \emph{worst-case scenario}, that is, in the case where variances of some random effects included in the (maximal) model are actually zero.}
In the following, we will show that there is a substantial cost: The additional protection against Type I errors implies a significant increase in Type II error rate or, in other words, a loss in statistical power to detect the significance of fixed effects. We will also show that selection of a parsimonious LMM is a promising alternative to the maximal model, balancing the Type I error rate and power.

\section{Simulation} \label{sec:simulation}
If there is any situation where the maximal model approach implies a cost in terms of statistical power, we should be able to demonstrate the problem with a simulation of a simple experiment, estimating the Grand Mean (intercept) and a single fixed effect as described below.

\subsection{Specification of the simulation models}
Here, $Y_{c,s,i}$ refers to the dependent variable, the subscripts stand for condition ($c$), subject ($s$) and item ($i$), the by-subject and by-item intercepts are $S_{0,s}$ and $I_{0,i}$, and the corresponding slopes are $S_{1,s}$ and $I_{1,i}$. The fixed effects are $\beta_{0}$ and $\beta_{1}$ and the residual error is $\varepsilon_{c,s,i}$.

\begin{align}
 Y_{c,s,i} &= \beta_{0}+S_{0,s}+I_{0,i}+\left(\beta_{1}+S_{1,s}+I_{1,i}\right)\, X_c + \varepsilon_{c,s,i}\,, \nonumber \\
 \left(\begin{array}{c} S_{0,s}\\ S_{1,s}\end{array}\right) &\sim 
   \mathcal{N} \left(\left(\begin{array}{c}0\\0\end{array}\right),
   \left(\begin{array}{cc} 
    \tau^2_{00} & \rho_S\tau_{00}\tau_{11}  \nonumber \\
    \rho_S\tau_{00}\tau_{11} & \tau^2_{11} \end{array}\right) \right)\,, \\
 \left(\begin{array}{c} I_{0,i}\\ I_{1,i}\end{array}\right) &\sim 
   \mathcal{N} \left(\left(\begin{array}{c}0\\0\end{array}\right),
   \left(\begin{array}{cc} 
    \omega^2_{00} & \rho_I\omega_{00}\omega_{11} \\
    \rho_I\omega_{00}\omega_{11} & \omega^2_{11} \end{array}\right) \right)\,, \nonumber \\
 \epsilon_{c,s,i} &\sim \mathcal(0, \sigma^2)\,. \nonumber
\end{align}

This generating process mimics a very simple experiment where $20$ items ($i=1,\dots,20$) are presented to $50$ subjects ($s=1,\dots,50$) under two conditions ($c=1,2$), encoded as $X_{c=1}=-0.5$ for the first and $X_{c=2}=+0.5$ for the second condition. For example, we can collect response times under two conditions with a grand mean of $2000$ ms and an experimental effect of $25$ ms. Accordingly, the model intercept was chosen as $\beta_{0}=2000$ and the experimental effect (i.e., the difference between the two experimental conditions) is either $\beta_{1}=0$ (assuming the H0) or $\beta_{1}=25$ (assuming H1). The key concern is how the complexity of the random-effects structure of the LMM affects the estimate and the significance of the fixed effect $\beta_{1}$, with respect to both Type I error and power.

In the generating process, there is a random intercept for each subject $S_{0,s}$ and a random slope (i.e., experimental effect) for the condition within each subject $S_{1,s}$. The standard deviation of the subject specific random intercept is $\tau_{00}=100$, while the standard deviation of the subject specific random slopes will be varied on the interval $\tau_{11}\in[0,120]$, to simulate the effect of the size of the random slope on Type I error rate and power. These two subject specific random effects are chosen to be correlated with $\rho_S=0.6$. 

Additionally there is a random intercept $I_{0,i}$ and slope $I_{1,i}$ for item. Again, the standard deviation of the random intercept is ($\omega_{00}=100$). Like the standard deviation of the subject specific random slope, the standard deviation of the item random slope will be varied on the interval $\omega_{11}\in[0,120]$ and the item specific random effects are again correlated with $\rho_I=0.6$. 

Finally, the model residuals are independent and identically distributed, $\varepsilon_{c,s,i}\sim\mathcal{N}(0,300^{2})$. Note that the item- and subject-related random slope standard deviations and the number of items and subjects were chosen to ensure that these variance components will be present, but barely detectable in the simulated data over a wide range of the interval $[1,120]$.

Given that every item is presented to each subject in each of the two conditions, the total number of data points is $2000$. From this generating process, a sample is drawn and five LMMs are fitted to the data that differ only in the structure of the random-effects part. The models are estimated under the null hypothesis of a zero fixed effect and under the alternative hypothesis of a fixed effect with value $\beta_1=25$.

The first model is the maximal model, including the estimation of correlation parameters ($\rho_I$ and $\rho_S$) that by definition are fixed at $0.6$ in the generating process. This model matches the generating process exactly unless the variances of the random slopes are set to 0.

\begin{align}
 Y_{c,i,s} &=  \beta_{0}+S_{0,s}+I_{0,i}+\left(\beta_{1}+S_{1,s}+I_{1,i}\right)\, X_c + \epsilon_{c,s,i}\,, \label{eq:max} \\
 \left(\begin{array}{c} S_{0,s}\\ S_{1,s}\end{array}\right) &\sim 
   \mathcal{N} \left(\left(\begin{array}{c}0\\0\end{array}\right),
   \left(\begin{array}{cc} 
    \tau^2_{00} & \rho_S\tau_{00}\tau_{11}  \nonumber \\
    \rho_S\tau_{00}\tau_{11} & \tau^2_{11} \end{array}\right) \right)\,, \\
 \left(\begin{array}{c} I_{0,i}\\ I_{1,i}\end{array}\right) &\sim 
   \mathcal{N} \left(\left(\begin{array}{c}0\\0\end{array}\right),
   \left(\begin{array}{cc} 
    \omega^2_{00} & \rho_I\omega_{00}\omega_{11} \\
    \rho_I\omega_{00}\omega_{11} & \omega^2_{11} \end{array}\right) \right)\,, \nonumber \\
 \epsilon_{c,s,i} &\sim \mathcal(0, \sigma^2)\,. \nonumber
\end{align}

Compared to the maximal one, the second model only differs in the two correlation parameters (i.e., it is model (Eq. \ref{eq:max}) where $\rho_S$ and $\rho_I$ are set to $0$).

\begin{align}
 Y_{c,i,s} &=  \beta_{0}+S_{0,s}+I_{0,i}+\left(\beta_{1}+S_{1,s}+I_{1,i}\right)\, X_c + \epsilon_{c,s,i}\,, \label{eq:id} \\
 \left(\begin{array}{c} S_{0,s}\\ S_{1,s}\end{array}\right) &\sim 
   \mathcal{N} \left(\left(\begin{array}{c}0\\0\end{array}\right),
   \left(\begin{array}{cc} 
    \tau^2_{00} & 0 \\
    0 & \tau^2_{11} \end{array}\right) \right)\,, \nonumber \\
 \left(\begin{array}{c} I_{0,i}\\ I_{1,i}\end{array}\right) &\sim 
   \mathcal{N} \left(\left(\begin{array}{c}0\\0\end{array}\right),
   \left(\begin{array}{cc} 
    \omega^2_{00} & 0 \\
    0 & \omega^2_{11} \end{array}\right) \right)\,, \nonumber \\
 \epsilon_{c,s,i} &\sim \mathcal(0, \sigma^2)\,. \nonumber
\end{align}

The third model is a reduced one which ignores the item specific random slope (i.e., it is model (Eq. \ref{eq:id}) where $\omega_{11}=0$).

\begin{align}
 Y_{c,i,s} &=  \beta_{0}+S_{0,s}+I_{0,i}+\left(\beta_{1}+S_{1,s}\right)\, X_c + \epsilon_{c,s,i}\,, \label{eq:woIS} \\
 \left(\begin{array}{c} S_{0,s}\\ S_{1,s}\end{array}\right) &\sim 
   \mathcal{N} \left(\left(\begin{array}{c}0\\0\end{array}\right),
   \left(\begin{array}{cc} 
    \tau^2_{00} & 0 \\
    0 & \tau^2_{11} \end{array}\right) \right)\,, \nonumber \\
 I_{0,i} &\sim \mathcal{N}\left(0,\omega_{00}^2\right)\,, \nonumber \\
 \epsilon_{c,s,i} &\sim \mathcal(0, \sigma^2)\,. \nonumber
\end{align}

The fourth model excludes the random slope for subject (i.e., it is model (Eq. \ref{eq:id}) where $\tau_{11}=0$) while keeping the random slope for item.

\begin{align}
 Y_{c,i,s} &=  \beta_{0}+S_{0,s}+I_{0,i}+\left(\beta_{1}+I_{1,i}\right)\, X_c + \epsilon_{c,s,i}\,, \label{eq:woSS} \\
 S_{0,s} &\sim \mathcal{N}\left(0,\tau_{00}^2\right)\,, \nonumber \\
 \left(\begin{array}{c} I_{0,i}\\ I_{1,i}\end{array}\right) &\sim 
   \mathcal{N} \left(\left(\begin{array}{c}0\\0\end{array}\right),
   \left(\begin{array}{cc} 
    \omega^2_{00} & 0 \\
    0 & \omega^2_{11} \end{array}\right) \right)\,, \nonumber \\
 \epsilon_{c,s,i} &\sim \mathcal(0, \sigma^2)\,. \nonumber
\end{align}

The fifth model excludes both random slopes (i.e., it is model (Eq. \ref{eq:id}) where $\tau_{11}=\omega_{11}=0$).

\begin{align}
 Y_{c,i,s} &=  \beta_{0}+S_{0,s}+I_{0,i}+\beta_{1}\, X_c + \epsilon_{c,s,i}\,, \label{eq:woSSIS} \\
 S_{0,s} &\sim \mathcal{N}\left(0,\tau_{00}^2\right)\,, \nonumber \\
 I_{0,i} &\sim \mathcal{N}\left(0,\omega_{00}^2\right)\,, \nonumber \\
 \epsilon_{c,s,i} &\sim \mathcal(0, \sigma^2)\,. \nonumber
\end{align}

\subsection{Two simulation scenarios}
There are two simulation scenarios. First, an appropriate model should yield adequate test statistics even if covariance parameters in the true model are set to zero. Obviously, these parameters would be removed in a reduced model (unless they were explicitly expected on theoretical grounds). In the first scenario, we set the effect-related (co-)variance parameters to zero; that is, we know that the maximal model is overparameterized with respect to these parameters. This scenario illustrates the \emph{maximal cost} of the maximal model specification.

Second, an appropriate model should not be affected by small (co-)variance components. Although they are present in the generating process, they are at or below the threshold of detectability. In this second scenario, we are simulating the situation where the maximal model actually matches the generating process, but with a relatively small sample size, this maximal random-effect structure may not be supported by the data. In such cases, as far as we know, it is unknown whether a model with a richer random-effect structure will outperform a parsimonious model without these variance components.

\subsection{Determination of the error rates} \label{sec:errorrates}
Both simulation scenarios involve two runs. In the first run, the Type I error rate of the models were estimated. In each iteration of the simulation, a sample was drawn from the generating process above, where $\beta_{1}=0$ (no fixed effect of condition in the generating process). Then all five models were fitted \citep[using the \emph{lme4} package version 1.1-13, ][]{Bates2015a,RCoreTeam2014} to this data excluding the fixed effect for condition (H0) and the same models where fitted to the same data including the fixed effect of condition (H1). If any of these ten model fits (including the maximal ones) did not converge, the sample was to be redrawn. Thus, it is ensured that the simulation results are not influenced by numerical convergence issues.

Then, the false-positive detection of the fixed effect for condition was determined with a likelihood ratio test (LRT). If the difference of the deviances between a model under H0 (excluding a fixed effect for condition) and the same model under H1 (including that fixed effect) is larger than $3.85$, the fit was considered as a false-positive detection of that fixed effect. This criterion implies a nominal Type I error rate of about $\alpha=0.05$, assuming a chi-squared distribution with one degree of freedom for the difference of the model deviances. In the second run, we determined power by drawing samples from the generating process with $\beta_{1}=25$ using the criterion $\chi^2_{1}\geq 3.85$ for the same models.

Given the Type I error rate and power estimates, the performance of the maximal model specification can be compared with the performance of the parsimonious model (see below). Obviously, the best model is the one providing maximal power while maintaining a Type I error rate at the nominal level (e.g., $\alpha=0.05$). The latter, however, not only depends on the model structure but also on the hypothesis test being used. For example, the LRT approximates the sample distribution of the deviance differences by a chi-squared distribution. This approximation immediately implies that the test statistic of an LRT for a fixed effect in an LMM is not exact and the obtained Type I error rate will not match the expected $\alpha$ exactly. This discrepancy between the expected and the observed Type I error-rate decreases with increasing sample sizes. The unknown exact Type I error rate of the test statistic, however, increases the difficulty of comparing the model performances. 

\subsection{Model selection}
With our simulations, we investigate the selection of a parsimonious, that is, a possibly reduced model for a given data set. Hence a criterion must be chosen to decide whether the complexity of a certain model is supported by the data. Naturally, the most complex model will always provide the best fit for a given data set, but bearing the risk of over-fitting the data. Therefore, every model selection criterion will try to balance the \emph{goodness-of-fit} with the model complexity.

Popular model selection criteria are the Akaike information criterion \citep[AIC, ][]{Akaike1998}, Bayes or Schwarz information criterion \citep[BIC, ][]{Schwarz1978} and the aforementioned likelihood ratio test. For our simulations, we use the LRT criterion (an evaluation of the AIC can be found in the Appendix). The BIC is known to put a strong penalty on the model complexity for small sample sizes \citep[e.g., ][]{Vandekerckhove2015}.

In contrast to the AIC or BIC, which allows us to compare several models at once, the LRT can only compare two models. Hence, an additional heuristic is needed to choose the \emph{best} model out of a set of candidate models. For our simulations, we chose the backward-selection heuristic. There, one starts with the most complex model (i.e., the maximal model, Eq. \ref{eq:max}) and reduces the model complexity (i.e., model 1 $\rightarrow$ 2 $\rightarrow$ 3 $\rightarrow$ 4 $\rightarrow$ 5) until a further reduction would imply a \emph{significant loss} in the goodness-of-fit. The \emph{significance level} of this model-selection criterion is specified by the chosen $\alpha_{LRT}$ of the LRT. Within the context of model selection, it is important to resist the reflex of choosing $\alpha_{LRT}=0.05$. The $\alpha_{LRT}$ cannot be interpreted as the "expected model-selection Type I error-rate" but rather as the relative weight of model complexity and goodness-of-fit. For example, choosing $\alpha_{LRT}=0$, an infinite penalty on the model complexity is implied and consequently the minimal model is always chosen as the \emph{best}, irrespective of the evidence provided by the data. Choosing $\alpha_{LRT}=1$ implies an infinite penalty on the goodness-of-fit, and the maximal model is always chosen as the \emph{best}. Therefore, choosing $\alpha_{LRT}=0.05$ may imply an overly strong penalty on the model complexity and hence select a reduced model even if data favor a more complex one. For our simulations, we chose $\alpha_{LRT}=0.2$.

In fact, when comparing two nested models, the LRT with $\alpha_{LRT}\approx 0.157$ is equivalent to the AIC. Hence, one may expect that the LRT with $\alpha_{LRT}=0.2$ puts a slightly larger penalty on the goodness-of-fit compared to the AIC and therefore, it will choose the more complex model more frequently (see also Appendix).

\section{Results}
The simulations were carried under two scenarios. In the first scenario, we investigated the cost incurred by the maximal model with respect to the power to detect the fixed effect. In this case, the variances of both random slopes where fixed to $0$ in the generating process (and, by implication, correlation parameters were zero as well). In this \emph{worst case scenario} for the maximal model, the intercept-only model (Eq.\ \ref{eq:woSSIS}) matches the generating process; all other models (Eqs. \ref{eq:max}-\ref{eq:woSS}) were overparametrized by construction.

In the second scenario, we varied the standard deviations of the random effect slopes from $\tau_{11}=\omega_{11}=0$ to $\tau_{11}=\omega_{11}=120$ while keeping the correlation of the within-subject random effects and the within-item random effects fixed at $0.6$. The goal was to determine how Type I error rates and power change as random slopes increase. Obviously, the maximal model should be favored for large random slopes. Moreover, the inclusion of models two to four in the simulations allowed us to examine whether model selection with a standard criterion yields better results than the maximal model when weak random slopes are present in the generating process.

\subsection{Worst case scenario}
The \emph{worst case scenario} allowed us to determine the maximal cost implied by the maximal model. In this case, the variance of both random slopes was set to zero in the generating process. Hence the intercept-only model (Eq.\ \ref{eq:woSSIS}) matched the generating process and the other four models were overparametrized.

Table \ref{tab:pathologic} summarizes the Type I error rate and power estimates, obtained in the simulation with 10,000 iterations for each model along with their $95\%$ confidence intervals (in parenthesis). Power decreased with model complexity. Moreover, the Type I error rates of all models were not substantially larger than the expected level $\alpha=0.05$. Unsurprisingly, model (Eq. \ref{eq:woSSIS}) provides the best power while maintaining a Type I error rate close to the nominal $\alpha=0.05$. Please note that test statistics of the LRT for the fixed-effect slope ($\beta_1$) is only approximately a $\chi^2$ distribution, even in the case where the model matches the generating process exactly. Hence, the chosen significance criterion does not correspond exactly to the expected $\alpha=0.05$. 

Unsurprisingly, the intercept-only model (Eq.\ \ref{eq:woSSIS}) rather than the maximal model (Eq.\ \ref{eq:max}) performs best for this \emph{worst case scenario}. Most interestingly, the maximal model has even a substantially smaller Type I error rate than the nominal $\alpha=0.05$. Hence it appears to be over-conservative at the cost of power. Similar results with a similar simulation were reported in \cite{Barr2013}. They, however, concluded that loss in power is negligible.

\begin{table}[h!t]
 \centering
 \begin{tabular}{| l |c|c|} \hline
   & Type I error rate & Power (1 - Type II error rate) \\ \hline \hline
  Model 1: Maximal & 0.0304 (0.0272, 0.0340)& 0.364 (0.354, 0.373)\\ \hline
  Model 2: Zero Correlation & 0.0331 (0.0386, 0.0368) &  0.377 (0.367, 0.386)\\ \hline
  Model 3: Zero Item Slope Var.\ & 0.0424 (0.0386, 0.0466) & 0.427 (0.417, 0.437)\\ \hline
  Model 4: Zero Subj Slope Var.\ & 0.0396 (0.0359, 0.0436) & 0.403 (0.393, 0.413)\\ \hline
  \textbf{Model 5: Random Intercepts} & 0.0510 (0.0468, 0.0555) & 0.455 (0.445, 0.465)\\ \hline
 \end{tabular}
 \caption{Type I error rates and power of models (Eqs. \ref{eq:max}-\ref{eq:woSSIS}) in the case of a generating process without random slopes; 95\% confidence intervals are also presented. In this case, random-intercept-only model (Eq. \ref{eq:woSSIS}) matches the generating process, while models (Eqs. \ref{eq:max}-\ref{eq:woSS}) are overparametrized by construction.
 } \label{tab:pathologic}
\end{table}

\subsection{Small random slopes}
In the first simulation scenario, we showed that the maximal model leads to substantially reduced power over the intercept-only model when the latter model matches the generating process. Hence, one may expect that even if the maximal model is true, that is, even in the case where the random slope variances of the generating process are non-zero, a reduced model (Eqs.\ \ref{eq:id}-\ref{eq:woSSIS}) may hold a power advantage over the maximal model (Eq.\ \ref{eq:max}), if its complexity is not supported by the data.

Of course, we do not know \emph{a priori} whether a certain random-effect structure is supported by the data. We can determine a parsimonious model with a standard criterion such as the LRT. This criterion weights the goodness-of-fit of each model with its complexity in order to select the \emph{best} model for a given data set and guard against over-fitting the data.

Therefore, in the second simulation scenario, we compared the performance (Type I error rates and power) of the maximal model (Eq.\ \ref{eq:max}) with the model selected by the LRT (out of Eqs.\ \ref{eq:max}-\ref{eq:woSSIS}) as a function of the random-slope variances \cite[see also][]{Westfall2014}. This scenario allowed us to study whether LRT detects the need for an increased model complexity with increasing random-slope standard deviations and therefore maintain a Type I error rate close to the nominal $\alpha$.

For each iteration of this simulation we chose the random-slope standard deviations in 20,000 steps from $\tau_{11}=\omega_{11}=0$ to $\tau_{11}=\omega_{11}=120$ (leading to a step size of $0.006$). All other variance-covariance parameters were kept constant. Along with the false-positive and false-negative detections, we obtained the Deviance values for each model. Then, we chose the best of the models \ref{eq:max}-\ref{eq:woSSIS} supported by the data according to the LRT model selection criterion and compared their performance. 

The false-positive and false-negative detections of each iteration are samples of a Bernoulli distribution with an unknown rate parameter (the Type I and II error rates, respectively). To visualize Type I error rate and power (1 - Type II error rate) as a function of the random-slope standard deviation, we fitted generalized additive models \citep[GAM, e.g.,][]{Hastie1990, Wood2006} to the false-positive and false-negative detection data using the binomial family to describe the response distributions. This not only accurately models the response distribution of the false-positive and false-negative detections, it also provides confidence intervals for the estimated Type I error rate and power.

\begin{figure*}[!ht]
 \centering
 \includegraphics[width=0.75\textwidth]{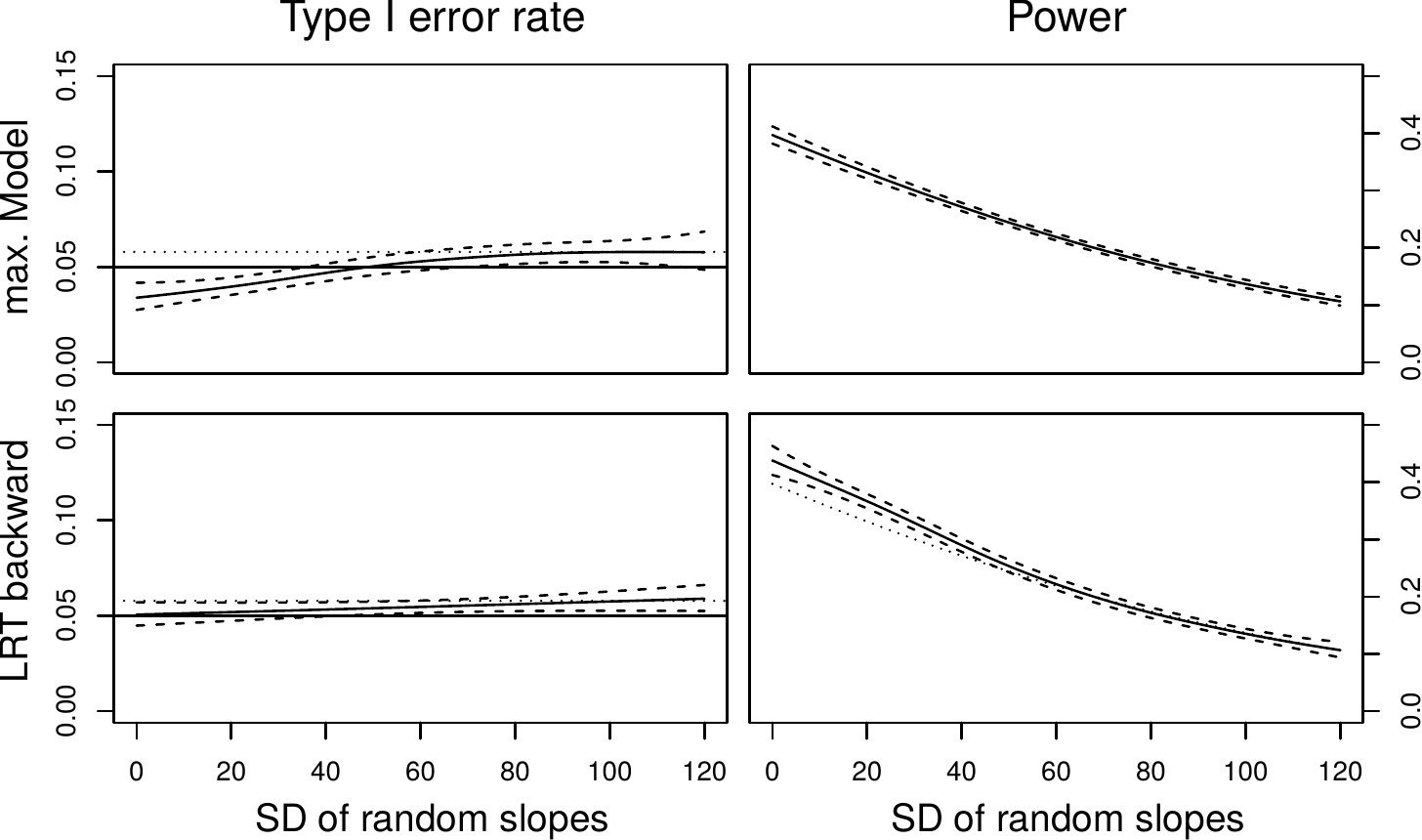}
 \caption{Comparison of Type I error rate (left panels) and power (right panels) for the maximal model (top panels) and the model selected according to LRT backward selection heuristic (bottom panels). The nominal Type I error rate ($\alpha=0.05$) is shown as a horizontal line in the left panels; the maximum Type I error rate of the maximal model is shown as a dotted line in the top left panel. The dotted line in the bottom right panel reproduces the power of the maximal model shown in the top right panel.} \label{fig:scanLRT1}
\end{figure*}

Figure \ref{fig:scanLRT1} shows Type I error rate and power (left and right column, respectively) as a function of the random-slope standard deviations ($\tau_{11}$, $\omega_{11}$) for the maximal model (Eq.\ \ref{eq:max}) and the model selected according to LRT.

The top row shows Type I error rate and power for the maximal model (Eq.\ \ref{eq:max}). Its Type I error rate appears to be substantially smaller than the expected $\alpha=0.05$ (horizontal solid line) up to $\tau_{11}=\omega_{11}\approx 40$. This indicates an over-conservative behavior of the maximal model in cases where the random slope standard deviations are too small to support the complex random-effect structure. On the interval $[80,120]$, the maximal model appears to be slightly anti-conservative with respect to the expected $\alpha=0.05$. On that interval, however, the maximal model matches the generating process.

Given that the maximal model is over-conservative even with respect to the expected $\alpha=0.05$ for small random slopes and, therefore, has reduced power, we expect that a parsimonious model may provide better power in these cases by choosing a model that is supported by the data. For example model (\ref{eq:woSSIS}) for small, model (\ref{eq:max}) for larger random slopes, and one of the models (\ref{eq:id}-\ref{eq:woSS}) in between.

The bottom row in Figure \ref{fig:scanLRT1} shows Type I error rate and power of the parsimonious model, selected according to LRT, as a function of the random-slope standard deviation. Although the Type I error rate appears to be larger than the expected $\alpha=0.05$, it is not substantially larger than the maximal Type I error rate of the maximal model. At the same time, it provides substantially better power on the interval $[0,40]$. Thus, the LRT-based model selection approach yields an advantage of statistical power over the maximal model while maintaining a comparable Type I error rate.

Not everyone may consider the gain in power for this setting (shown in Fig. \ref{fig:scanLRT1}) as \emph{substantial enough} to justify the additional effort one has to put into model selection. However, the relative power gain naturally increases when overall power decreases (e.g., when sample size is small). Hence, we performed the same simulation with a reduced number of subjects ($30$ instead of $50$) and a reduced number of items ($10$ instead of $20$). To this end we reduced the total sample size from $2000$ to $600$.

\begin{figure*}[!ht]
 \centering
 \includegraphics[width=0.75\textwidth]{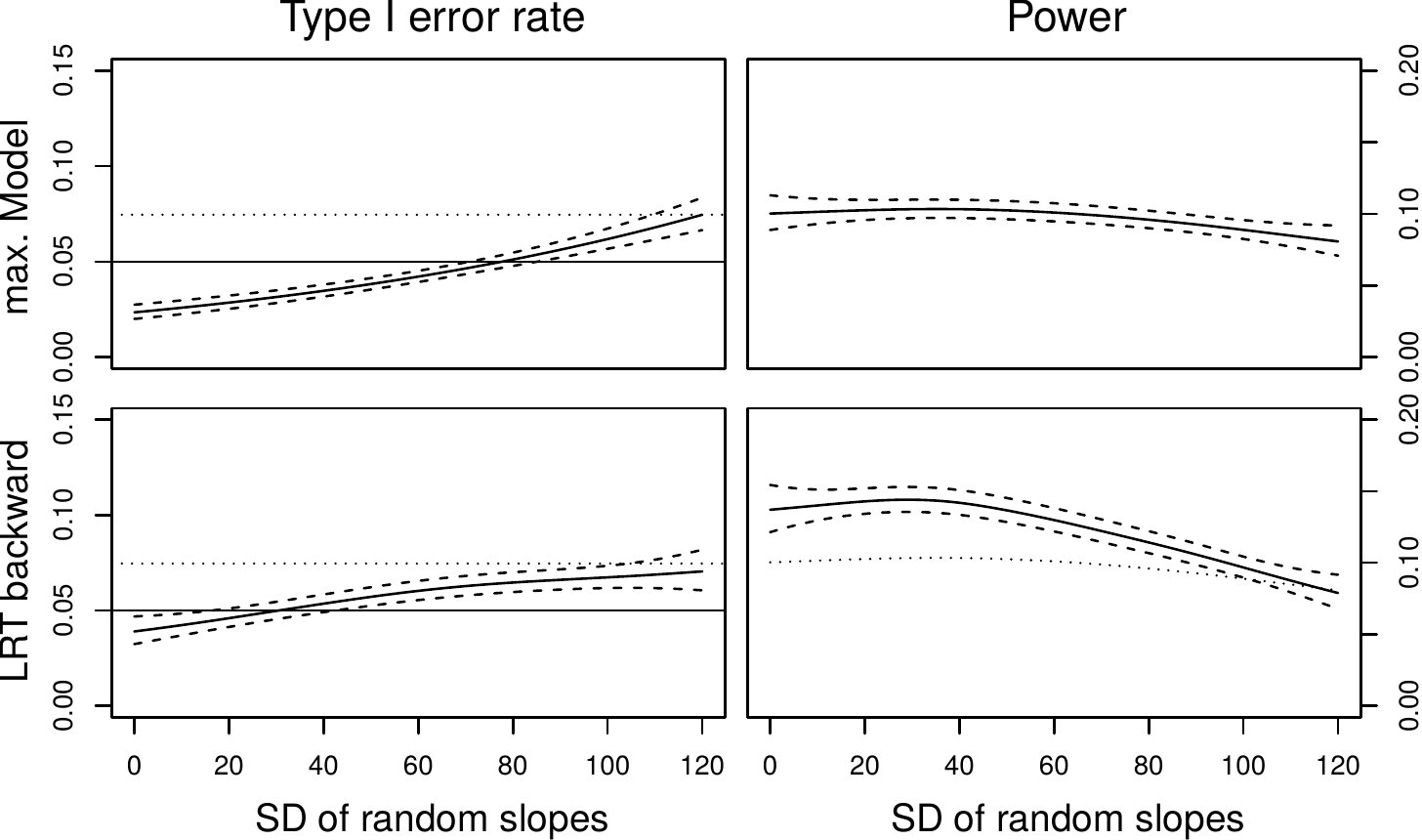}
 \caption{Comparison of the Type I error rate (left panels) and power for the maximal model (top panels) and the model selected according to LRT backward selection heuristic (bottom panels) for a smaller sample size. The nominal Type I error rate ($\alpha=0.05$) is shown as a horizontal line in the left panels; the maximum Type I error rate of the maximal model is reproduced as a dotted line in the top left panel. The dotted line in the bottom right panel reproduces the power of the maximal model shown in the top right panel.} \label{fig:scanLRT2}
\end{figure*}

Figure \ref{fig:scanLRT2} shows the Type I error rates and power for the maximal model (top panels) and the model selected according to LRT (bottom panels) for a smaller sample size ($N_{subj}=30, N_{item}=10$). Here the maximal model shows an anti-conservative behavior on the interval $[0,80]$. Consequently, a model selection approach using the LRT backward selection heuristic is able to provide a substantial gain in power over a larger interval $[0,90]$ while maintaining a Type I error rate comparable with the maximal model. Moreover, the relative power gain is larger for this reduced sample size.

\begin{figure*}[!ht]
 \centering
 \includegraphics[width=0.5\textwidth]{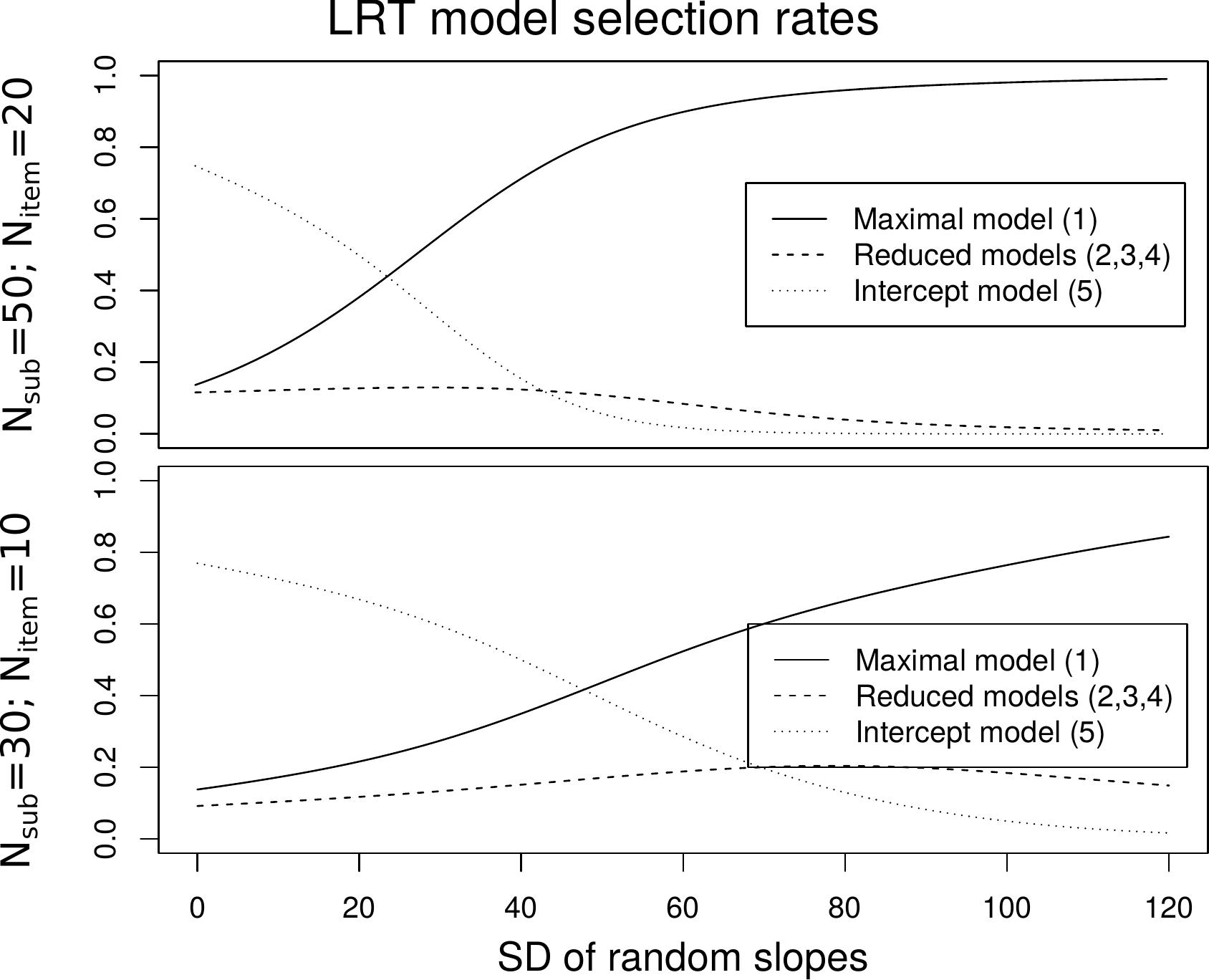}
 \caption{Comparison of the model selection rates by the LRT backward selection heuristic for a large ($N_{subj}=50;N_{item}=20$, top panel) and a smaller sample size ($N_{subj}=30;N_{item}=10$, bottom panel), as a function of the random-slope standard deviation for the maximum model (solid line), a reduced model (Eqs. \ref{eq:id}-\ref{eq:woSS}, dashed line) and the random-intercept only model (dotted line).} \label{fig:modsel}
\end{figure*}

Figure \ref{fig:modsel} shows how the LRT model selection rate changes with increasing random-slope variances (top panel: $N_{subj}=50; N_{item}=20$, bottom panel: $N_{subj}=30; N_{item}=10$). A similar pattern is obtained for both sample sizes: The intercept-only model \emph{wins} for small random-slope standard deviations and is less frequently selected in favor of the more complex models with increasing random-slope standard deviations; the maximal model gets selected almost exclusively for large random effect slopes. 

These model selection rates also explain the behavior seen in Figures \ref{fig:scanLRT1} and \ref{fig:scanLRT2}: Small random slopes are not supported by the data, even in the case of a larger sample size. For this case, a parsimonious model yields the best description of the data and provides a power advantage over a maximal model. As the random-slope standard deviations increase, a more complex model must be chosen to describe the data adequately. Of course, this happens earlier for larger sample sizes compared to smaller sample sizes. Hence, a model selection approach using the LRT backward selection heuristic provides a power benefit compared to the maximal model.

In summary, the first simulation scenario (i.e., the simulation of the worst-case scenario) revealed clear deficits of the maximal model. In the second simulation scenario, even when the generating process matched the maximal model, the maximal model performed worse than a parsimonious model when variance components were not well supported by the data.

\section{Discussion}
The simulations yielded a clear set of results. First, in agreement with \citet{Barr2013} and others before them, we showed that the maximal model is able to guard against an increased Type I error rate by ignoring a significant variance component. However, while the maximal model indeed performs well as far as Type I error rates were concerned, power decreases substantially with model complexity. We have shown that the maximal model may trade-off power for some conservatism \emph{beyond} the nominal Type I error rate, even in cases where the maximal model matches the generating process exactly. In fact, the best model is the one providing the \emph{largest power}, while maintaining the chosen nominal Type I error rate. If more conservatism with respect to the Type I error rate is required, the significance criterion $\alpha$ should be chosen to be more conservative, instead of choosing a possibly over-conservative method with some unknown Type I error rate. 

As already stated by \citet[p. 185]{stroup2012generalized}: 
\begin{quotation}
Neither the [maximal] nor the [minimal] linear mixed models are appropriate for most repeated measures analysis. Using the [maximal] model is generally wasteful and costly in terms of statistical power for testing hypotheses. On the other hand, the [minimal] model fails to account for nontrivial correlation among repeated measurements. This results in inflated [T]ype I error rates when non-negligible correlation does in fact exist. We can usually find middle ground, a covariance model that adequately accounts for correlation but is more parsimonious than the [maximal] model. Doing so allows us full control over [T]ype I error rates without needlessly sacrificing power.
\end{quotation}

Our simulations have shown that determing a parsimonious model with a standard model selection criterion is a defensible choice to find this middle ground between Type I error rate and power (see also Appendix). Such a model maintains a Type I error rate similar to the maximal model at a better statistical power. Experimental designs, as currently employed in psycholinguistic research, are likely to be compatible with parsimonious models. A survey of $15$ studies on Chinese relative clauses \citep{VasishthetalPLoSOne2013} and of $26$ of the studies reviewed by \citet{JaegerEngelmannVasishth2016} showed that the number of subjects used in rating studies and reading studies (self-paced reading and eyetracking) range from $16$ to (in one case) $150$ \citep{JaegerEngelmannVasishth2015}, and the number of items from $12$ to $80$; numbers at the lower ends of the range are far more common for subjects and items in psycholinguistics. For such typical sample sizes, it is not necessarily the maximal, but more likely a model with a parsimonious random effect structure that will be most suitable for describing the data of factorial experiments. For experiments with much larger sample sizes (relatively high power), the situation could be different.

There is a looming concern about using model selection in inferential statistics about fixed effects in factorial experiments. Factorial experiments typically implement designs with a limited number of balanced conditions. Usually, full-factorial ANOVAs are specified providing test statistics for all fixed effects, that is, all main effects and interactions. Rarely, main effects or interaction terms are pooled with error terms because there is no \emph{a priori} expectation for them to be significant. Model selection is almost never used to pool non-significant main effects and interaction terms because they are not significant. \citet{Barr2013} refer to the first two cases as confirmatory and the last case as exploratory hypothesis testing. How do (or how should) these options influence decisions about the choice of the random-effect structure of LMMs? \citeauthor{Barr2013} provide a nuanced discussion of these issues and carefully delineate various alternatives, but in the end they come down strongly in favor of their recommendation \emph{to keep it maximal}, that is against model selection.\footnote{Unfortunately, their nuanced discussion did not register with the field, and the title of the paper, \textit{keep it maximal} became the main take-away point, which began to be enforced by journals and reviewers.} In line with our results, we want to make a case for model selection.

\citet{Barr2013} contrast confirmatory (design-driven; i.e., the LMM is selected before the analysis and does not depend on the data--unless there are convergence problems) and data-driven hypothesis testing (i.e., the final LMM is selected taking into account whether model parameters are supported by the data). If we assume that all within-subject or within-item fixed effects are expected to be different from zero, it is reasonable to assume that all variance components are significantly different from zero as well, because it is unlikely that our experiments will detect a natural constant in psycholinguistic experiments. From this perspective, the maximal model looks like a coherent analysis strategy. This is only very rarely a realistic scenario, at least once we go beyond a two-factorial design; we simply don't see explicit expectations about the pattern of means relating to three-factor interactions. Confirmatory hypothesis testing must pool terms that are not expected to be significant and their associated variance components with the residual variance. Thus, strictly speaking, traditional ANOVA-based hypothesis testing is more often better characterized as overfitting rather than confirmatory. \citeauthor{Barr2013} favor the maximal model because they assume that removal of non-significant terms might be the consequence not of \emph{a priori} selection of a model, but a post-hoc selection with an eye towards the significance of p-values in the fixed effects. 

Our starting point is different. The goal of model selection is not to obtain a significant p-value; the goal is to identify the most parsimonious model that can be assumed to have generated the data. Strictly speaking, a purely confirmatory hypothesis testing approach of only ever fitting a single maximal model and publishing the resulting p-values would also require that we never check model assumptions (such as normality of residuals). Since violation of model assumptions can lead to serious misinterpretation of the data, the researcher has no choice but to engage in model criticism, which requires a certain amount of exploratory modeling, as discussed by \citet{gelmanhill07}. 
The researcher who rigorously wants to carry out confirmatory hypothesis testing may be better off following the advice of \citet{deGroot2014} to clearly separate the data that is used for exploratory analysis from the data used for confirmatory analysis. Fitting maximal models will in any case not immunize the researcher from the danger of researcher degrees of freedom.

There are four arguments in favor of a model-selection strategy. First, as shown with the simulations in this article, fixed-effect estimates do not depend on random-effects structure as long as the factorial design is balanced (i.e., covariates are uncorrelated). Thus, the invariance of fixed-effect estimates in balanced designs for varying random-effects structures is not a surprise; one may just as well use ordinary least squares to estimate them. The random-effects structure may impact the estimate of the standard errors associated with fixed effects and thereby at least sometimes (not always) the decision about statistical inference. Matters get more complicated once one deals with correlated covariates, non-factorial designs, substantial imbalance due to missing data, or auto-correlated residuals \citep{Baayen2016Cave, Matuschek2015}. Our simulations showed that, in the long run, the parsimonious model yields the best chances to detect a true fixed effect as significant. 

Second, while plausible, the assumption that the true value of all variance components is larger than zero is actually not the critical test during model selection. Rather the question is for which variance components this assumption is supported by the data. Fitting the maximal model incurs substantial loss of statistical power if the true value of the variance component is small. Moreover, such overparameterized LMMs often yield nonsensical estimates of correlation parameters (i.e., values of -1 or +1) that clutter journal pages and mislead readers.

Third, we strongly want to encourage a move beyond the interpretation of variance components as mere nuisance parameters that need to be taken care of, for example, to avoid anti-conservative estimates of fixed effects \cite[e.g.,][]{Barr2013, Schielzeth2009}. They hold the key for a joint consideration of experimental effects and associated individual or item differences \citep{Gonzalez2014,KlieglEtAl2011}. For example, whether a correlation between mean reponse time and an experimental effect is estimated as reliably positive or negative may have profound consequences for a theoretical consideration. Also, a fixed effect may not be significantly different from zero, but model selection may reveal reliable individual differences in this effect \citep{KlieglEtAl2011}. In this case, the significant variance component points to qualitative differences in the direction of effects (i.e., positive and negative for about 50\% of the subjects, respectively).

Fourth, the distinction between design-driven and data-driven, as introduced by \citet{Barr2013}, misses an important confirmatory aspect in multivariate statistics: Any hypothesis about the support of variance components by the data requires a model comparison. For example, hypotheses about the random-effect structure (e.g., postulating a correlation parameter to be significantly different from zero) are tested by model selection (i.e., an LRT of models with and without the critical variance component). In multivariate statistics, LRTs are confirmatory, but \citeauthor{Barr2013} call this analysis data-driven because selection of the parsimonious model depends on the data. Finally, as \citeauthor{Barr2013} also point out, model-selection strategies must be carefully documented.

Having reached this conclusion, we face the discussion about suitable model selection criteria (e.g., AIC, LRT, GCV) and efficient schemes (e.g., forward, backward, full-scale cross validation), like all other scientific fields using multivariate statistics to guide the accumulation of knowledge \citep[e.g.,][]{Vandekerckhove2015}. There are also some specific proposals for selection of parsimonious mixed models \citep[e.g.,][]{Bates2015}.  The question of which model selection approach is should be chosen is relevant because we want to know which approach will maximize power. However, the choice of model selection method may be less important if we run experiments with higher power in the first place. 
In contrast to many other fields, psycholinguistic research enjoys two key advantages. First, it is usually not difficult to design experiments with sufficient statistical power; that is, to plan \emph{a priori} for a larger number of subjects and number of items (although in some cases it can be more difficult to increase items than subjects). Aiming for high power is important for independent reasons, because even a statistically significant result obtained with a low-power experiment could easily have the wrong sign or an exaggerated effect size \citep{gelman2014beyond,colquhoun2014investigation}. Second, it is not too difficult to repeat an experiment with new samples of subjects and items to validate the results of the \emph{best} model found on the basis of the data of the first experiment. These are definitely viable and reliable steps towards determining the best random-effects structure for the data of a specific design.

Finally, we want to emphasize that we are not proposing a new dogma that is an alternative to the ``keep it maximal'' proposal of \citet{Barr2013}. We do not insist  that now everyone must engage in model selection. There is certainly room for argument about different approaches here, as there is in all of statistical practice. In order to achieve maximum transparency in the analysis, we propose that all data and code should be released with the publication of a paper, so that readers can revisit the analysis and investigate the robustness of claimed effects. This allows for checks of, for example, whether the exclusion of a random effect was justified or a maximal model should have been specified. Even if the maximal model converges, researchers may have valid reasons for determining a parsimonious model, especially if they want to use all of the statistical power their research program affords in the long run.

\section{Acknowledgements}
This research was supported by Deutsche Forschungsgemeinschaft as part of the Research Group 868  \emph{Computational Modeling of Behavioral, Cognitive, and Neural Dynamics}. The simulation R source code is available at \emph{Potsdam Mind Research Repository}, \texttt{http://read.psych.uni-potsdam.de/}. We are grateful to the reviewers for their helpful comments on earlier versions of this paper.

\bibliographystyle{model5-names}
\bibliography{references}

\clearpage
\appendix
\section*{Appendix: Model selection using the AIC} \label{app:modelsel}
Throughout the main article, we employed the LRT backward heuristic as the model selection criterion. Another popular means of model selection is the Akaike information criterion \citep[AIC; ][]{Akaike1998}. In contrast to the LRT, the AIC allows for the comparison of several models at once, while the former only compares two models with each other. The simultaneous comparison of several models avoids the need to choose a heuristic for a sequential, pair-wise model selection. Within this appendix, we present the results (Type I error rates and power) of our simulations using the AIC as the model selection criterion instead of the LRT.

\begin{figure*}[!ht]
 \centering
 \includegraphics[width=0.75\textwidth]{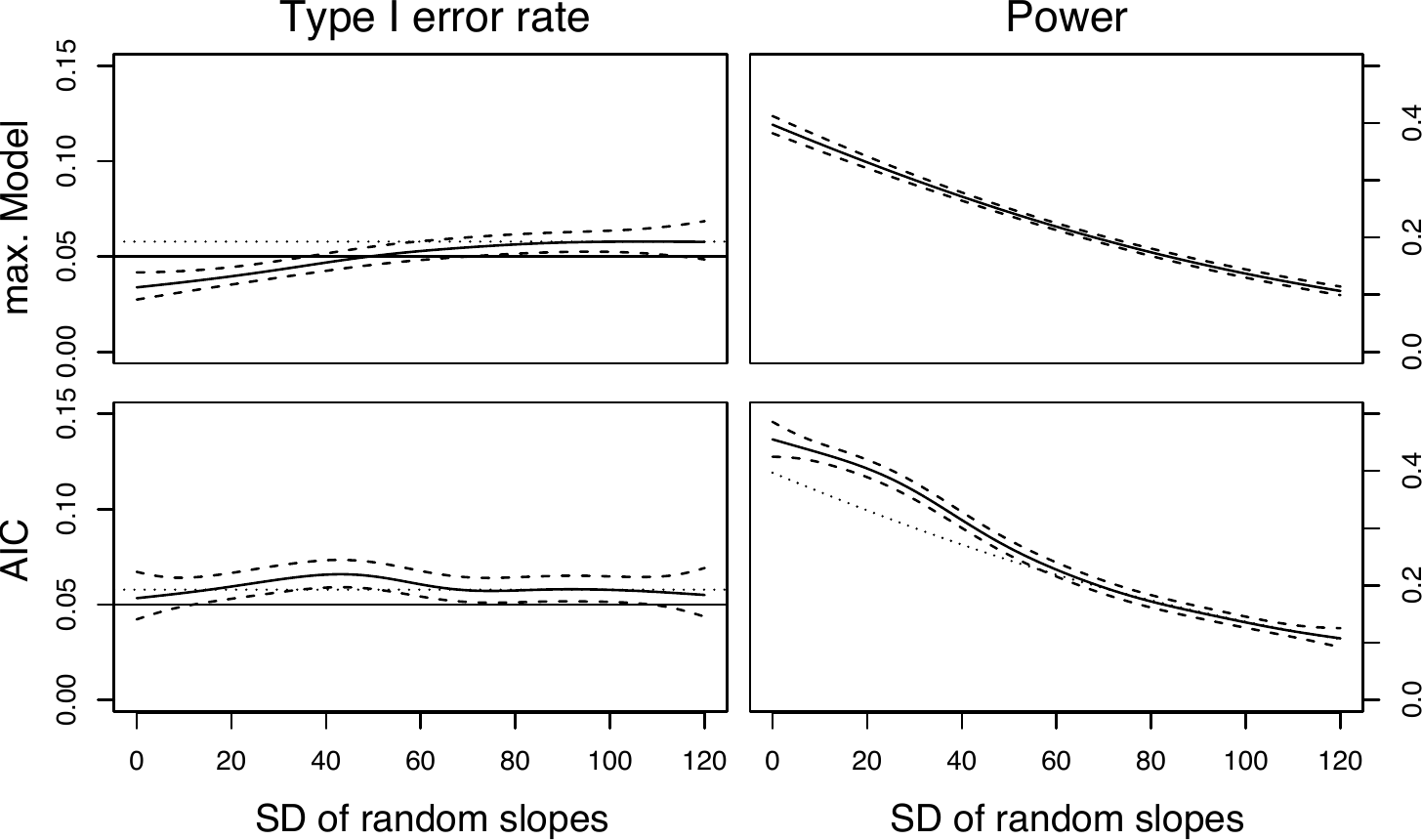}
 \caption{Comparison of the Type I error rate (left panels) and power for the maximal model (top panels) and the model selected according to AIC (bottom panels). The nominal Type I error rate ($\alpha=0.05$) is shown as a horizontal line in the left panels; the maximum Type I error rate of the maximal model is reproduced as a dotted line in the top left panel. The dotted line in the bottom right panel reproduces the power of the maximal model shown in the top right panel.} \label{fig:scanAIC1}
\end{figure*}

\begin{figure*}[!ht]
 \centering
 \includegraphics[width=0.75\textwidth]{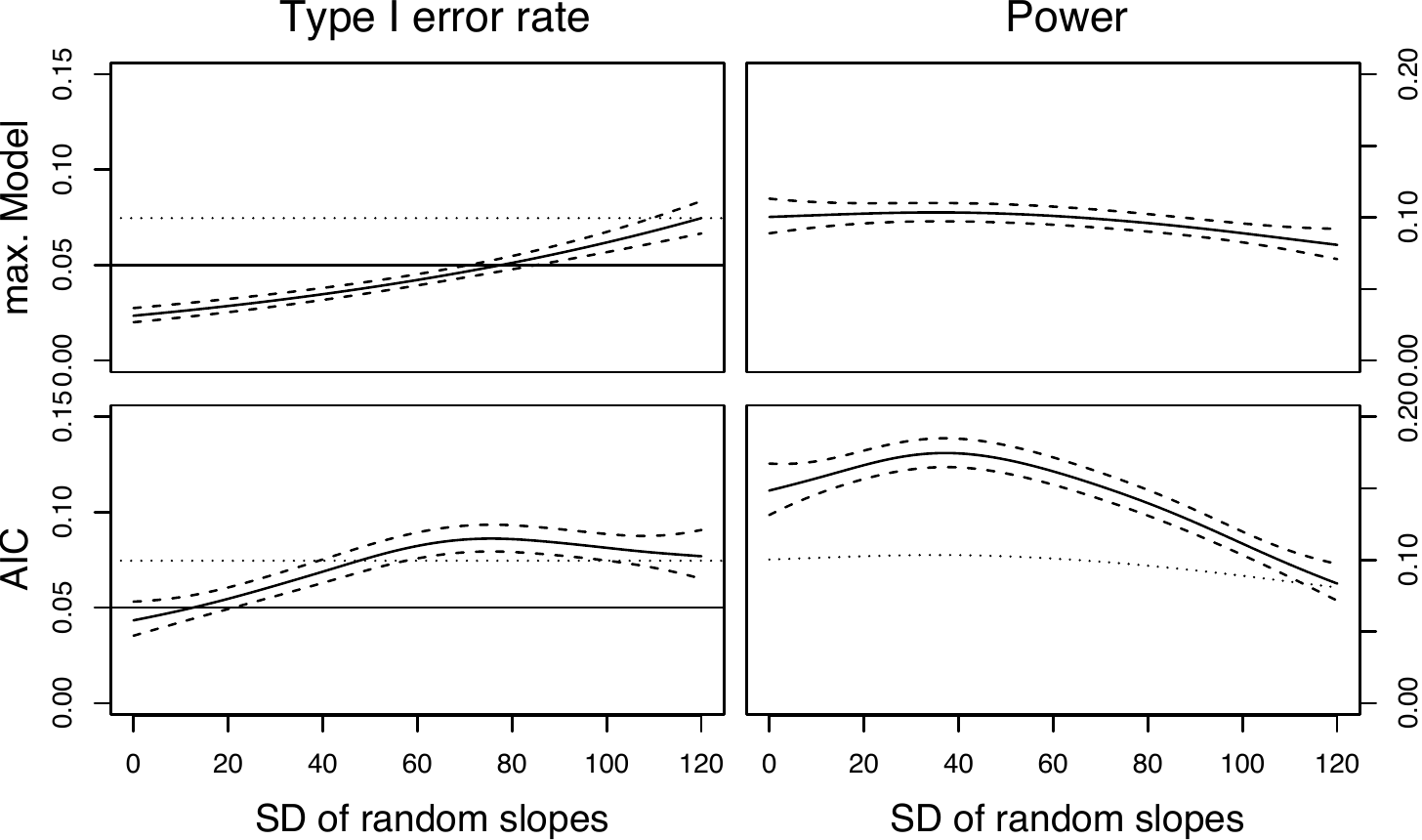}
 \caption{Comparison of the Type I error rate (left panels) and power for the maximal model (top panels) and the model selected according to AIC (bottom panels) for a small sample size. The nominal Type I error rate ($\alpha=0.05$) is shown as a horizontal line in the left panels; the maximum Type I error rate of the maximal model is reproduced as a dotted line in the top left panel. The dotted line in the bottom right panel reproduces the power of the maximal model shown in the top right panel.} \label{fig:scanAIC2}
\end{figure*}

Figure \ref{fig:scanAIC1} shows the Type I error rate (left column) and power (right column) of the maximal model (top row, identical to the top row of Figure \ref{fig:scanLRT1}) and the model selected by the AIC (bottom row) for a larger sample size (i.e., $N_{sub}=50$, $N_{item}=20$). Like the LRT, the AIC provides a clear power benefit over the maximal model while maintaining a comparable Type I error rate (compare Fig. \ref{fig:scanLRT1}). For a smaller sample size (i.e., $N_{sub}=30$, $N_{item}=10$, Fig. \ref{fig:scanAIC2}), however, the model selected by the AIC might be considered as too anti-conservative with respect to the Type I error rate, although a definite statement is not possible as the exact Type I error rate of the test is unknown.

The difference between the two model selection approaches for the smaller sample size (compare Fig. \ref{fig:scanLRT2} and \ref{fig:scanAIC2}) is explained by the larger weight on the goodness-of-fit of the LRT with $\alpha_{LRT}=0.2$ compared to the AIC. The backward selection scheme prevents one from choosing a simpler model if only a small amount of evidence is provided by the data in favor of it. Meaning, the scheme \emph{stays with the more complex model} if the selection criterion is indefinite, particularly in cases of small samples sizes.

In general, model selection tries to balance the goodness-of-fit of a model with its risk for overfitting the data. Therefore, the decision about which model is the most appropriate one for a given data set depends strongly on the amount of evidence provided by the data. Simply taking the \emph{best} according to the model selection criterion in cases of very small data sets bears the risk of choosing the \emph{wrong} model. To this end, model selection gets increasingly difficult with decreasing samples sizes while allowing for an increased gain in (relative) power.

In summary, for small sample sizes, the model selected by the AIC appears to be slightly anti-conservative even with respect to the maximum Type I error rate of the maximal model. The more conservative LRT with backward-selection heuristic, however, still maintains a Type I error rate near the one of the maximum model, while providing a substantial benefit in power compared to the maximal model (compare Figs. \ref{fig:scanLRT2}, \ref{fig:scanAIC2}).

\end{document}